\documentstyle[preprint,aps]{revtex}
\begin{document}
\draft
\preprint{SISSA.Ref.70/97/CM}

\title{
QUANTUM COHERENT ATOMIC TUNNELING BETWEEN TWO 
TRAPPED BOSE-EINSTEIN CONDENSATES 
}
\author{A. Smerzi $^{1}$, S. Fantoni $^{1,2}$,
S. Giovanazzi$^{1}$ and S. R. Shenoy$^{2}$
}

\address{ {1)} International School of Advanced Studies,  
via Beirut 2/4, I-34014, Trieste, Italy } 
\address{ {2)} International Centre for Theoretical Physics,  
I-34100, Trieste, Italy } 
%\date{\today}
\maketitle
%\\[3pt]
\begin{abstract}
%\\ \medskip}\author{\small\parbox{14cm}{\small
We study the coherent atomic tunneling
between two zero-temperature Bose-Einstein condensates (BEC) 
confined in a double-well magnetic trap. Two Gross-Pitaevskii equations 
for the self-interacting BEC amplitudes, coupled by a transfer 
matrix element, describe the dynamics in terms of the inter-well
phase-difference and population imbalance. In addition to
the anharmonic generalization 
of the familiar $ac$ Josephson effect and plasma 
oscillations occurring in superconductor junctions, 
the non-linear BEC tunneling dynamics sustains
a self-maintained population imbalance: a novel "macroscopic
quantum self-trapping effect".
%\vspace{.5in}
\end{abstract}
{
%{\pacs{SISSA.ref.169/96/CM/MB}}
%}}\address{}\maketitle

\pagebreak
\narrowtext

The recent 
experimental observation of the Bose-Einstein condensation (BEC) in a
dilute gas of 
trapped atoms \cite{1,2}, 
has generated much interest
in the properties of this new state of matter.
A quite fascinating possibility  
is the observation of new quantum phenomena on macroscopic scales,
related with the 
superfluid nature of the 
condensate. In fact, broken symmetry arguments show that 
the condensate atoms can be described by a common,
"macroscopic" one-body wave function
$\Psi(\vec r,t) = \sqrt{\rho}~ e^{i \theta}$ (the order parameter),
with $\rho$ the condensate density.
For a weakly interacting BEC, 
the order parameter obeys a 
non-linear Schroedinger, or
Gross-Pitaevskii equation (GPE) \cite{3}:
$$
i \hbar {{\partial \Psi}\over {\partial t}} =
- {\hbar^2 \over{2 m}}\nabla^2 \Psi +
[ V_{ext}(\vec r)  + g_0 \| \Psi \|^2] \Psi
\eqno(1)$$
where $V_{ext}$ is the external potential and 
$g_0 = {{4 \pi \hbar^2 a}\over m}$ is the inter-atomic 
scattering pseudo potential, with 
$a,m$ the atomic scattering length and mass respectively.

The GPE has been successfully applied to investigate the 
collective mode frequencies of a trapped BEC in the linear
regime \cite{4}, the relaxation times of monopolar oscillations 
\cite{5}, 
and, because of the non-linear self-interaction, it could also induce chaotic 
behavior in dynamical quantum observables \cite{5,6}

The existence of a macroscopic quantum phase (difference) was 
dramatically demonstrated recently \cite{2}. 
A far off-resonant intense laser sheet divided a 
trapped condensate, creating a large barrier in between. 
Switching off the double-well trap, the two released condensates
 overlapped,
producing a robust "two-slit" atomic interference pattern, 
clear signature of
phase coherence over a macroscopic scale $(\simeq 10^{-2} cm)$.
The $non$ destructive detection of phase-differences between 
two trapped BEC could be achieved by lowering the 
intensity of the laser sheet. This allows for atomic tunneling through the 
barrier, and the detection of Josephson-like current-phase
effects \cite{2,7,8,9}. 
In superconductor Josephson junctions (SJJ), 
phase coherence signatures 
include a $dc$ external voltage producing an $ac$ current, 
or the "plasma" 
oscillations of an
initial charge imbalance \cite{10,11}. 
For neutral superfluid He II, 
voltage drives, tunnel junctions, or capacitive charges are absent. The
only accessible Josephson analogue \cite{12} involves two He II baths 
connected by 
a sub-micron orifice, at which vortex phase-slips \cite{13} support a chemical
potential (height) difference, through 
the Josephson frequency relation.

Although the trapped BEC is also a neutral-atom  Bose system,
its population can be monitored by phase-contrast microscopy;
the double-well curvatures and barrier heights can be tailored by the
position and the intensity of the laser sheet partitioning the magnetic
trap \cite{2}. 
We note that the chemical 
potential between the two condensates depends both on
the zero-point energy difference, that acts like an external "dc" SJJ
voltage, and on the non-linear interaction that, through 
an initial 
population imbalance, acts like a capacitive SJJ charging energy. 
Thus, we suggest that 
the BEC tunnel junction can show the analogues of the familiar 
Josephson effects in supercunductor junctions, with the ability to
tailor traps and the atomic interaction compensating for electrical 
neutrality.

In this Letter we study the atomic tunneling at zero temperature between 
two non-ideal, weakly linked BEC
in a (possibly asymmetric) double well trap. This induces
a coherent, oscillating flux of atoms between wells, that is a signature
of the $macroscopic~superposition~of~states$ in which the condensates
evolves.
The dynamics 
is governed by two Gross-Pitaevskii equations for the BEC amplitudes, 
coupled by a transfer matrix element (Josephson tunneling term). 
Analogues of the superconductor Josephson effects such as the $ac$ effects 
and "plasma" oscillations are predicted. We also find that the non-linearity
of the dynamic tunneling equations produces a novel 
self-trapping effect. 

Consider a double-well magnetic trap 1,2 as in Fig.(1).
This system can be described by a two-state model: 
$$
i \hbar {{\partial \psi_1}\over {\partial t}} =
(E^0_1 + U_1 N_1) \psi_1 - K \psi_2 
\eqno(2a)$$
$$
i \hbar {{\partial \psi_2}\over {\partial t}} =
(E^0_2 + U_2 N_2) \psi_2 - K \psi_1 
\eqno(2b)$$
with uniform amplitudes 
$\psi_{1,2} = \sqrt{N_{1,2}}~ e^{i \theta_{1,2}}$, where 
$N_{1,2}$, $\theta_{1,2}$ are the number of particles and phases
in the trap $1,2$ respectively, and $K$ is the coupling matrix element 
\cite{14}.
The parameters $E^0_{1,2},~U_{1,2}$ and $K$ can be determined from 
appropriate overlap integrals of the time-independent GPE eigenfunctions
$\Phi_{1,2}$ of the isolated traps, as outlined later. The total number of 
atoms $N_T = N_1 + N_2$ is constant, but we stress that a coherent 
phase description, i.e. the existence of 
definite phases $\theta_{1,2}$, implies that the phase fluctuations 
($\simeq {1 \over \sqrt{N_{1,2}}}$ \cite{15})
must be small, giving $N_{1,2} > N_{min} \simeq 10^3$, say.

We note that the BEC inter-trap tunneling Eq.s(2) are similar 
in form to models of single-trap atomic-level transitions \cite{8}, 
or polaron 
hopping in semiclassical approximation \cite{16}, although they 
describe quite different physics.

In terms of the phase-difference $\phi = \theta_2 - \theta_1$ and fractional 
population difference $-1 < z = {{N_1 - N_2}\over N_T} < 1$, Eq.s(2) become
$(\hbar = 1)$: 
$$ \dot{z} = - \sqrt{1 - z^2}  sin \phi \eqno(3a)$$
 $$ \dot{\phi} = \Lambda z + {z \over \sqrt{(1 - z^2)}} cos \phi + \Delta E
\eqno(3b)$$
where the time has been rescaled as 
$2 K t \to t$. 
The dimensionless parameters are:
$$\Delta E = {(E^0_1 - E^0_2)/(2 K) + (U_1 - U_2) N_T/(4 K)} \eqno(4b)$$
$$\Lambda = (U_1 + U_2) N_T/(4 K) \eqno(4c)$$

For two symmetric traps, $E^0_1 = E^0_2$ ($\Delta E = 0$), 
$U_1 = U_2 = U$, and
$\Lambda = U N_T /2 K$.
In the following  we will assume a positive scattering
length $a$ ($\Lambda > 0$); note, however, that Eq.s(3) are
invariant under the trasformation 
$\Lambda \to - \Lambda,~\phi \to -\phi + \pi,~\Delta E \to \Delta E$.

The $z$, $\phi$ variables are canonically conjugate, with 
$
\dot{z} = - {{\partial H} \over {\partial \phi}}
$,
$
\dot{\phi} =  {{\partial H} \over {\partial z}}
$
and the Hamiltonian is given by:
$$
H = {\Lambda \over 2} z^2 - \sqrt{1-z^2}~ 
cos\phi + \Delta E~ z
\eqno(5) 
$$
In a simple mechanical analogy, $H$ describes a $non-rigid$ pendulum,
of tilt angle $\phi$ and length  proportional to $\sqrt{1-z^2}$,
that decreases with the "angular momentum" $z$.  

The inter-trap tunneling current is given by:
$$I = \dot{z} {{N_T} \over2} = I_0  \sqrt{(1 - z^2)} sin\phi;
~~~~~~ I_0 = K N_T \eqno(6)$$
The detailed 
analysis of Eq.s(3) with exact analytical solutions will be 
presented elsewhere \cite{17};  
here we consider
three regimes.

1)$Non-interacting~limit$.
For symmetric wells and negligible interatomic interactions
($\Lambda \to 0$), Eq.s(2) can be solved exactly, yielding Rabi-like
oscillations in the population of each trap with a frequency:
$$ \omega_R = 2 K \eqno(7)$$
as studied in \cite{7,8}. However,
the ideal Bose gas limit is not accessible experimentally. 

2)$Linear~regime$. 
In the linear limit $(|z| <<1, |\phi| << 1)$
Eq.s(3) become:
$$\dot{z} \simeq - \phi \eqno(8)$$
$$\dot{\phi} \simeq (\Lambda + 1) z \eqno(9)$$
These describe the small amplitude oscillations of the pendulum 
analogue, with a sinusoidal $z(t)$ with a frequency:
$$ \omega_L = \sqrt{2 U N_T K + 4 K^2} \eqno(10)$$ 
The BEC oscillations of population should show up 
as temporal oscillations of phase-contrast patterns \cite{2},
or other probes of atomic population \cite{1}. 

Linearizing Eq.s(3) in $z(t)$ only, we obtain:
$$
\dot{z} \simeq - sin \phi
\eqno(11a)$$
$$
\dot{\phi} \simeq \Delta E + (\Lambda + cos \phi) z \eqno(11b)$$
$$I \simeq I_0 sin \phi \eqno(11c)$$
For large trap asymmetries with 
$\Delta E  >> (\Lambda + cos(\phi) )z$, 
we have: $\phi = \phi(0) + \Delta E t$, 
giving an
oscillating $z(t)$ with frequency:
$$\omega_{ac} \simeq E^0_1 - E^0_2
\eqno(12)$$ 
where an "ac" current $I(t)$ is produced by the "dc" 
trap asymmetry $\Delta E$. 
It is simple to show that a small 
oscillation in the laser position, or in its intensity  
($K \to K (1 + \delta sin(\omega_0 t)),~ \delta <<1$), will result in a 
$dc$ inter-trap current of non-zero time average 
$<I(t)> \simeq \delta <sin(\omega_0 t) sin(\omega_{ac} t) > \ne 0$, at
resonant match $\omega_0  =  \omega_{ac}$.
This is the analogue of the Shapiro effect \cite{10}
 in superconductor junctions. 

In SJJ, the current of Cooper pairs 
$N_{1,2}$ is $I_{SJJ} = -2 e (\dot{N_1} - \dot{N_2}) = 2 e E_J~ sin\phi$,
and the Josephson frequency relation for the relative phase is 
$\dot{\phi} = \Delta \mu
= 2 e V + (N_1 - N_2) E_c$, for a 
$dc$ applied voltage $V$ and a junction capacitance $C$, 
with $E_c = (2 e)^2/2 C$ \cite{10}. 
These SJJ equations can be directly compared with the BEC Eq.s(3). It is 
then clear  
that the 
$ac$ Josephson frequency 
$\omega_{ac} = 2 e V$ 
and the Josephson "plasma"
frequency \cite{10,11} $\omega_p = \sqrt{E_c E_J}$ 
are the analogue of Eq.(12) and Eq.(10) respectively. 
Note however that, for
SJJ, $\omega_p$ is independent of the barrier cross section $A$,
since $E_J \simeq A$ and 
$E_c \simeq C^{-1} \simeq A^{-1}$,
while the BEC has $\omega_{L} \simeq A^{1/2}$
since $K \simeq A$, and the bulk energy $U N_T$ is approximately
A-independent.

3)$Non-linear~regime$.
A numerical solution of Eq.s(3) yields 
non-sinusoidal oscillations, that are
the anharmonic generalization of the Josephson effects. Moreover,
an additional 
novel non-linear effect occurs in the BEC: a self-locked 
population imbalance.

Fig.(2) shows solutions of Eq.s(3) with initial conditions $z(0) = 0.6$,
$\phi(0) = 0$ and illustrative parameters $\Lambda = 1,8,9.99,10,11$ 
respectively.
The sinusoidal oscillations around $z=0$ became anharmonic as 
$\Lambda$ increases, Fig.(2a,b), 
and with a precursor slowing down, Fig.(2c),
there is a critical transition for $\Lambda = \Lambda_c = 10$, dashed line in
Fig(2d). Then for $\Lambda = 11$ the population in each trap oscillates
around a non-zero  
time averaged $<z(t)> \ne 0$, solid line in Fig.(2d).
In the non-rigid pendulum analogy, 
this corresponds to an 
initial angular momentum $z(0)$ sufficiently large to swing the 
pendulum bob over the 
$\phi = \pi$ vertical orientation, with a non-zero $<z(t)>$
average angular momentum corresponding to the rotatory motion. This critical
behavior depends  
on $\Lambda_c = \Lambda_c [z(0), \phi(0)]$, as can be easily
found from the energy conservation constraint. In fact from Eq.(5), 
the value $z(t)= 0$ is inaccessible at any time  if:
$$
\Lambda > 2~ ({{\sqrt{1 - z(0)^2}~cos(\phi(0)) + 1} \over z(0)^2}) \eqno(13)$$
The full dynamical behavior of Eq.(3) is summarized
in Fig(3), that shows the $z-\phi$ phase portrait with constant energy lines.
There are energy minima along $z = 0$ at $2 n \pi$, and "running"
solutions $<\dot{\phi(t)}> \ne 0$ with $<z(t)> \ne 0$, moving along the 
sides of these wells. The vertical points $\phi = (2 n+ 1) \pi$, that would be
isolated unstable points for the rigid pendulum, now support oscillations of 
restricted range, as a consequence of non-rigidity, i.e. non-linearity.

The self-trapping of an initial BEC population imbalance, seen in 
Fig.(2d) and Fig.(3) arises because of the 
inter-atomic interaction in the Bose gas 
(non-linear self-interaction in GPE). It has a quantum nature, 
involving the
coherence of a macroscopic number of atoms. It differs from single polaron
trapping of an electron in a medium \cite{16} 
and from gravitational effects on 
He II baths \cite{12}. It can be considered as a novel
"macroscopic quantum self-trapping" (MQST).

Non-linear effects like MQST are unobservable in SJJ. The requirement that the 
chemical potential difference 
$\mu_1 - \mu_2 \simeq (N_1 - N_2) E_c$ must be less than the
quasiparticle gap $2 \Delta_{qp}$ (to prevent a jump off the Josephson I-V
branch) implies a stringent constraint in imbalances 
$|z | < (2 \Delta_{qp}/E_c N_T) \simeq 10^{-12}$ for typical 
parameters. For the BEC, the requirement that tunneling does not access 
excitation energies is much less restrictive. As an example, 
let consider two weakly linked condensates of $N_T \simeq 10^4$ atoms,
confined in two symmetric spherical traps with 
frequency $\omega_0 \simeq 100~Hz$. Evaluating Eq.s(17) below with a simple
variational wave function \cite{18},  
we have $E^0 \simeq 0.5~ nK$, $U N_T \simeq 3 ~nK$, and from 
an estimation of the excitation gap we obtain the constraint $|z| \le 0.5$. 
Taking $K \simeq 0.1~ nK$, we have 
$\Lambda \simeq 10$, close to the onset of the critical behavior.
Increasing the number of particles or the trap frequencies, 
the value of $\Lambda$ can increase, making MQST 
observable. Typical frequency oscillations are 
$\omega_L \simeq KHz$ for $N_T \simeq 10^6$, 
that should be compared with the plasma frequency of
SJJ that are \cite{11} of the order of $\omega_p \simeq~GHz$. 

We now outline the  
derivation of the parameters $E_0, U, K$.

To this purpose we note that in the barrier region the modulus of the 
order parameter in the GPE is exponentially small.
This allows us to look for an eigenstate of GPE 
of the form:
$$
\Psi = \psi_1(t) \Phi_1(x) +
\psi_2(t) \Phi_2(x)
\eqno(14)$$
where $\Phi_1, \Phi_2$ are the ground state solutions for isolated
traps with $N_1 = N_2 = N_T/2$.

Replacing Eq.(14) in the GPE Eq.(1)  
with the conditions: 
$$\int \Phi_1 \Phi_2 d \vec r \simeq 0 \eqno(15)$$
and
$$\int \|\Phi_1\|^2 d \vec r= \int  \| \Phi_2\|^2 d \vec r = 1
\eqno(16)$$
we obtain:
$$
E^0_{1,2} =  \int [{\hbar^2 \over{2 m}} |\nabla \Phi_{1,2}|^2
+ \Phi_{1,2}^2 V_{ext}]  d \vec r 
\eqno(17a)$$
$$
U_{1,2} = g_0 \int  \Phi_{1,2}^4 d \vec r  
\eqno(17b)$$
$$ 
K = - \int [{\hbar^2 \over{2 m}} (\nabla \Phi_1 \nabla \Phi_2) 
+ \Phi_1 \Phi_2 V_{ext}] d \vec r
\eqno(17c)$$

At finite temperature 
the interaction between the normal component of the
Bose gas with the condensate should be included, and the parameters become
temperature dependent. 
Such corrections are small for temperatures smaller than excitation energies.
High density
BEC could induce quasiparticle/collective mode scattering
with finite lifetime
of the coherent oscillations \cite{5}; phase diffusion 
could induce phase coherence
collapse and revival \cite{19}. These effects deserve further studies.
 
In conclusion, the BEC coherent atomic tunneling 
in a 
double-well trap induces
non-linear population oscillations that are a
generalization of the sinusoidal Josephson effects 
familiar in superconductors. A novel population imbalance occurs for
parameters beyond critical values:
a macroscopic quantum self-trapping effect.

\vskip 0.2truecm
Discussions with S.Raghavan and useful references from
L. Bonci and G. Williams are acknowledged.
%\pagebreak

\pagebreak

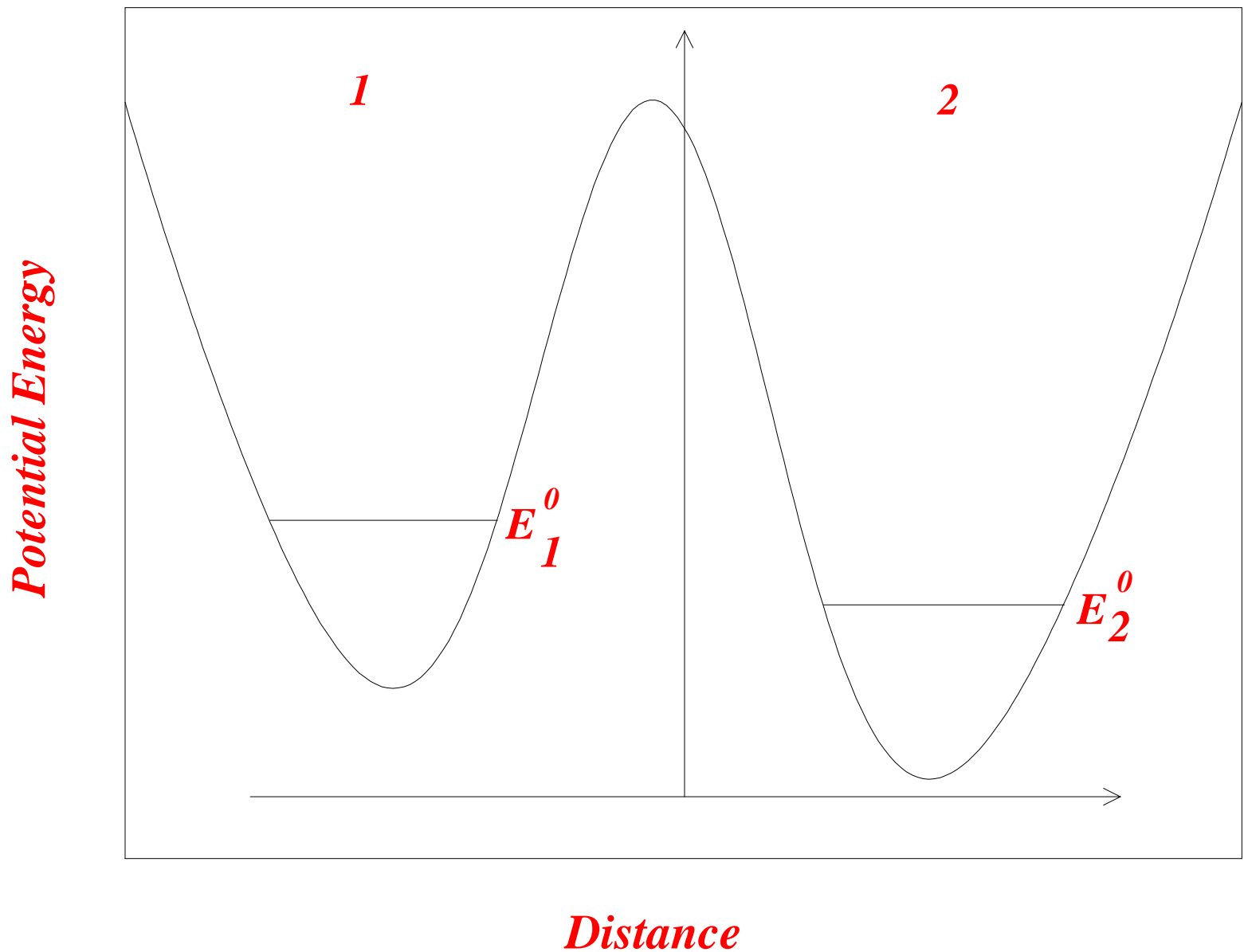
\begin{figure}
\caption{
The double well trap for two Bose-Einstein condensates with $N_{1,2}$ and
$E^0_{1,2}$ the number of particles and the zero-point energies in the 
trap $1,2$ respectively.}
\end{figure}
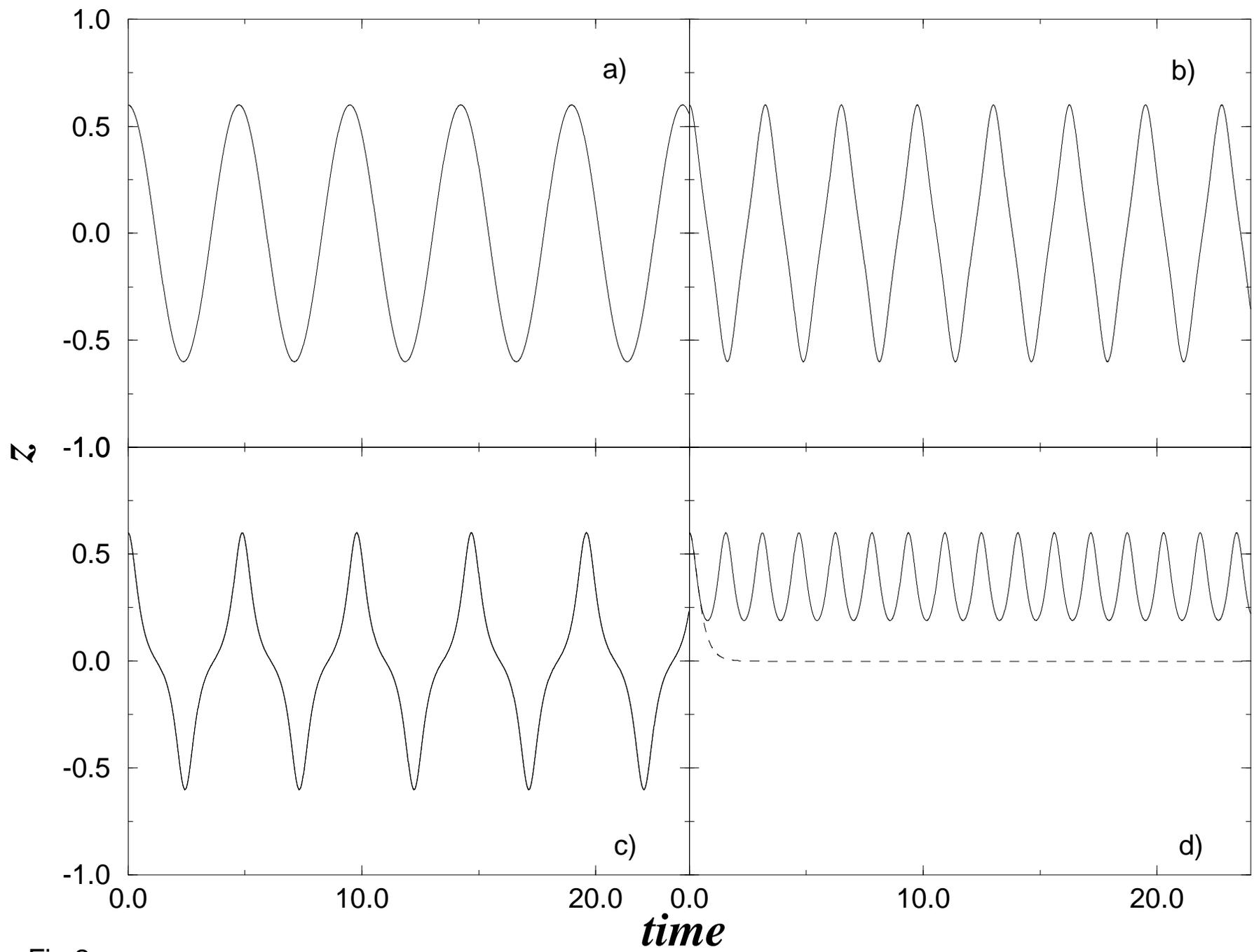
\begin{figure}
\caption{Fractional population imbalance $z(t)$ versus 
rescaled time, with initial conditions $z(0)=0.6$,
phase difference $\phi(0)=0$, and 
$\Lambda = 1$ (a), $\Lambda = 8$ (b), $\Lambda = 9.99$ (c),
$\Lambda = 10$ (dashed line, d), $\Lambda = 11$ (solid line, d).}
\end{figure}
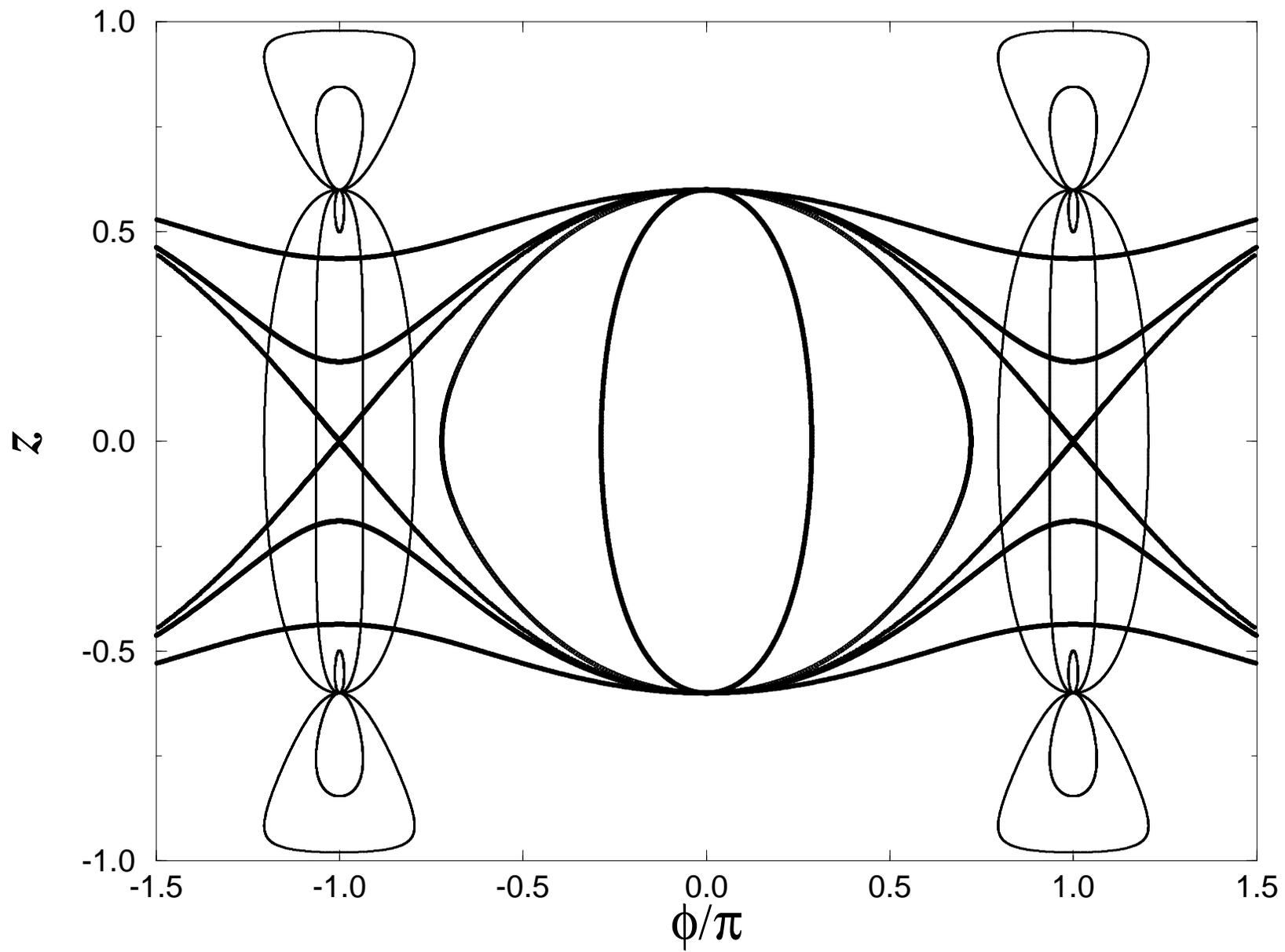
\begin{figure}
\caption{Constant energy lines in a phase-space plot 
of population imbalance $z$ versus phase difference $\phi$.
Bold solid-line: $z(0) = 0.6,~\phi(0)=0,~\Lambda = 1,8,10,11,20$.
Solid line: $z(0) = 0.6,~\phi(0) = \pi,~\Lambda = 0,1,1.2,1.5,2$.}
\end{figure}

\end{document}